\documentclass[preprint]{aastex631}

\usepackage{graphicx}

\usepackage{amsmath}

\begin{document}

\title{What if we find nothing? Bayesian analysis of the statistical information of null results in future exoplanet habitability and biosignature surveys}

\correspondingauthor{Daniel Angerhausen}
\email{dangerhau@ethz.ch}

\author[0000-0001-6138-8633]{Daniel Angerhausen}
\affiliation{ETH Zurich, Institute for Particle Physics \& Astrophysics, Wolfgang-Pauli-Str. 27, 8093 Zurich, Switzerland}
\affiliation{Blue Marble Space Institute of Science, Seattle, WA, USA}
\affiliation{SETI Institute, 189 N. Bernado Ave, Mountain View, CA 94043, USA}

\author[0000-0002-3929-6932]{Amedeo Balbi}
\affiliation{Dipartimento di Fisica, Università di Roma ``Tor Vergata'', Italy}

\author[0000-0001-5139-1978]{Andjelka B. Kova{\v c}evi{\'c}}
\affiliation{University of Belgrade-Faculty of Mathematics, Department of Astronomy, Studentski trg 16, 11000 Belgrade, Serbia}

\author[0000-0003-2530-9330]{Emily O. Garvin}
\affiliation{ETH Zurich, Institute for Particle Physics \& Astrophysics, Wolfgang-Pauli-Str. 27, 8093 Zurich, Switzerland}

\author[0000-0003-3829-7412]{Sascha P. Quanz}
\affiliation{ETH Zurich, Institute for Particle Physics \& Astrophysics, Wolfgang-Pauli-Str. 27, 8093 Zurich, Switzerland}
\affiliation{ETH Zurich, Department of Earth and Planetary Sciences, Sonneggstrasse 5, 8092 Zurich, Switzerland}



\begin{abstract}
Future telescopes will survey temperate, terrestrial exoplanets to estimate the frequency of habitable ($\eta_{\text{Hab}}$) or inhabited ($\eta_{\text{Life}}$) planets. This study aims to determine the minimum number of planets ($N$) required to draw statistically significant conclusions, particularly in the case of a null result (i.e., no detections). Using a Bayesian framework, we analyzed surveys of up to $N=100$ planets to infer the frequency of a binary observable feature ($\eta_{\text{obs}}$) after null results. Posterior best fits and upper limits were derived for various survey sizes and compared with predicted yields from missions like the Large Interferometer for Exoplanets (LIFE) and the Habitable Worlds Observatory (HWO). Our findings indicate that $N=20-50$ ``perfect'' observations (100\% confidence in detecting or excluding the feature) yield conclusions relatively independent of priors. To achieve 99.9\% upper limits of $\eta_{\text{obs}} \leq 0.2/0.1$, approximately $N \simeq 40/80$ observations are needed. For ``imperfect'' observations, uncertainties in interpretation and sample biases become limiting factors. We show that LIFE and HWO aim for sufficiently large survey sizes to provide statistically meaningful estimates of habitable environments and life prevalence under these assumptions. However, robust conclusions require careful sample selection and high-confidence detection or exclusion of features in each observation.

\end{abstract}



\section{Introduction} \label{sec:intro}
In the past two decades, the discovery of thousands of exoplanets has led to increasingly accurate estimates of the occurrence of rocky worlds in the habitable zone of their stars. At the same time, the focus of observations has been gradually shifting from  detection to characterization, with a number of current and future instruments and missions devoted to refining the sample of potentially habitable nearby planets and to perform detailed spectroscopic study of their atmospheres. These include the Transiting Exoplanet Survey Satellite \citep[TESS,][]{Ricker2014}, the CHaracterising ExOPlanets Satellite \citep[CHEOPS,][]{Benz2020}, PLAnetary Transits and Oscillations of stars \citep[PLATO,][]{Rauer2014} and Atmospheric Remote-sensing Infrared Exoplanet Large-survey \citep[ARIEL,][]{Tinetti2021} space missions, the ground-based European-led Extremely Large Telescope \citep[ELT,][]{2007Msngr.127...11G} and JWST  \citep[][]{Greene2016}, as well as future projects currently under study, such as the Habitable Worlds Observatory \citep[HWO,][]{2021pdaa.book.....N} or the Large Interferometer for Exoplanets \citep[LIFE,][]{LIFE1}.  The ambitious long-term goal of such observations is assessing the actual habitability of temperate, terrestrial exoplanets in our galactic neighborhood, and possibly detecting spectroscopic signatures of global biospheres \citep{Seager2014b,Grenfell2017,Fujii2018,Schwieterman2017}. 

One of the great accomplishments of exoplanets science in the past years was the derivation of first estimates of $\eta_{earth}$ -- the mean number of earth-like rocky planets per star -- using the large number of exoplanet detections made by the \textit{Kepler} \citep[][]{2010Sci...327..977B} and {TESS} satellites as well as ground based transit and radial velocity surveys \citep[see, e.g.,][]{2018ApJ...856..122K,Bryson2021,Bergsten2022}.
In analogy to this, one of the prospective outputs of the surveys discussed here, will be to derive estimates of the frequency of habitable $\eta_{\text{Hab}}$ or even inhabited planets $\eta_{\text{Life}}$. 

First empirical estimates for $\eta_{\text{Hab}}$ or $\eta_{\text{Life}}$ will certainly have an enormous impact. However, it is complicated to observationally constrain these quantities by remote detections of limited samples alone (see below and in the discussion in section \ref{conclusions}). These variables are among the major remaining unknowns essential for evaluating the prevalence of life in the universe. For example, they play a pivotal role in the renowned Drake equation and its modifications \citep{Drake1965, Seager2018}. Furthermore, these estimates have a direct impact on gauging the potential future lifespan of humanity, influencing assessments of possible bottlenecks, often referred to as the ``Great Filter'', in the evolution of life \citep{Haqq_Misra_2020,2016AsBio..16....7C}. For instance, the surveys discussed here could show that even on a planet with suitable conditions, abiogenesis might still be an extremely rare event. If these future observations on the other hand reveal that simple life is abundant, but complex or intelligent life is absent, it would suggest that subsequent evolutionary steps (e.g., multicellularity, intelligence, or technological advancement) are suppressed by the ``Great Filter''.

Unfortunately, our current knowledge is restricted only to the emergence of life on Earth, making it exceedingly challenging to draw meaningful inferences about the prevalence of life elsewhere. This limitation is compounded by the presence of selection biases, which render very problematic to make conclusive assessments \citep{Balbi2023}. Nevertheless, some limited insights can be gleaned from the relatively rapid appearance of life on our planet \citep{Spiegel2011, 2020PNAS..11711995K}.
In this context, it is important to highlight additional relevant insights and investigations that contribute to our understanding of planetary systems, the factors influencing habitability, and the broader search for extraterrestrial life.
Examples are planet formation models that predict the occurrence of habitable planets \citep[e.g.,][]{2021A&A...656A..70E, 2021A&A...656A..72B, 2022NatAs...6.1296K}, the `red sky paradox,' which questions why Earth orbits a G-type star rather than a later-type star \citep{2021PNAS..11826808K}, studies of the prevalence of planetary conditions that may support abiogenesis \citep[e.g.,][]{2014OLEB...44..339B,2016AsBio..16...68R,2020Life...10...52L,2020SciA....6.3419S,2022Life...12.1429D} and/or habitable conditions \citep[e.g.,][]{2018ApJ...864...75K, 2022NatAs...6..189K, 2024PSJ.....5....3S}, or formal birth-death relationships \citep{2024arXiv240707097K}.

Building on previous studies  such as \citet{Balbi2020a} and \citep{2022ExA....54.1197Q}, which analyzed the potential impact of future biosignature surveys on our understanding of the frequency of inhabited worlds in our galaxy, this work offers a novel perspective. 

We aim to estimate the minimum number of exoplanets required for future observations to yield statistically significant conclusions regarding $\eta_{\text{Hab}}$ and $\eta_{\text{Life}}$.  Our focus is specifically on the clearly defined scenario of a null result, i.e., where no surveyed planet shows signs of habitability or life. Without diverting into a detailed discussion, individual observations of the survey that would actually indicate a habitable planet could be the detection of ocean glint \citep{Robinson_2010,Lustig-Yaeger_2018}, strong heat redistribution (Fujii et. al., Braam et al., in prep) on a tidally locked planet or certain amounts of water vapor in the atmosphere (Rugheimer et al., in prep) that point at the presence of surface water. For the individual (unambiguous) detection of life on these planets it could be observations of (capstone) bio- \citep[e.g.,][]{LIFE12} or even technosignatures \citep[e.g.,][]{2024ApJ...969...20S}.  
Initially, we adopt an assumption of 'perfect knowledge' in our individual observations, implying absolute certainty in the absence of detected features, thus eliminating the possibility of false negatives or positives. This baseline helps us understand the upper limits of potential results of surveys, when being absolutely certain that each observation accurately reflects the true state of the planet. 
Similar methodological issues have been discussed in different contexts such as general statistics, particle physics, and high energy astrophysics \citep[see e.g.,][]{10.1214/ss/1009213286,2007ASPC..371...75C,2010ApJ...719..900K}.

Of course, this is a highly idealized scenario, as remarked in several earlier studies. For instance, \cite{Truitt_2020}, \cite{Catling_2018}, and \cite{Walker_2018} presented (Bayesian) frameworks to quantify the confidence in individual observations, while \cite{smith2023}, \cite{foote2022false}, and \cite{fisher2023complex} discussed the problems with false positives/negatives, underscoring that \textit{``high confidence in life detection claims require either (1) a strong prior hypothesis about the existence of life in a particular alien environment, or conversely, (2) signatures of life that are not susceptible to false positives.''} 
This highlights the importance of prior probabilities and robust likelihood functions.
Furthermore, \cite{vickers2023} addressed the problem of unconceived alternative explanations of otherwise agreed upon biosignatures by emphasizing the need to account for unknown hypotheses in the Bayesian model. \cite{Green_2021} and \cite{gillen2023} discussed solutions to better quantify and communicate the confidence in individual (non-)~detections using Bayesian credible intervals and posterior distributions. We attempt to take  these intricacies into account in section \ref{sec:met_imp} of our analysis, where we model the observational uncertainty by introducing a probability of obtaining false negative results or prior knowledge about a biased sample. 

\section{Method}

The main questions we want to address in our study are these: assuming we observed a number $N$ of planets in a given survey and can exclude a certain feature (i.e. habitability or the presence of life) for all of them, how well can we quantify the universal fraction $\eta_{\text{obs}}$ of planets having this feature (e.g., $\eta_{\text{Hab}}$ or $\eta_{\text{Life}}$)? And how does it impact our ability to derive these fractions if we are not 100\% confident in each single observation in our survey of this feature?

We tackle this problem by simulating a sample of $N$ individual observations, with $N$ varying between 1 and 100. This allows us to assess the significance of the results, by asking for which number $N$ of observed planets we can cross certain thresholds in our derived knowledge about the absolute fraction $\eta_{\text{obs}}$ having the desired feature.

We first adopt a Bayesian approach and use a set of assigned prior beliefs for $\eta_{\text{obs}}$ (``optimistic'', ``neutral'' or ``pessimistic'', see Section \ref{def_priors}) to compute the posterior belief distribution while conditioning on the findings of null results. The results from the posterior distribution are obtained over $N$ observations using Bayesian updating procedures. We use the posterior distribution to derive the credible intervals for $\eta_{\text{obs}}$ as a function of $N$, and identify the minimal sample size requirements for a given level of assigned posterior belief. We then also compare to a Frequentist approach to the same set of questions (see Section \ref{sec:met_freq}).

\subsection{The Beta-Binomial model}\label{bbmodel}

In this study we are dealing with observations asking a \textit{``yes or no''} question in each individual element of our survey. Therefore we adopt a Beta distribution to represent our preliminary understanding/prior belief $f(\eta_{\text{obs}})$ of the fraction of a binomially distributed observational parameter $\eta_{\text{obs}}$ (for example $\eta_{\text{Hab}}$ or $\eta_{\text{Life}}$) $\in [0, 1]$.
The Beta distribution is a suitable choice for modeling our prior distribution of $\eta_{\text{obs}}$ for several reasons. Firstly, it is defined on the interval [0, 1], aligning with the range of probabilities. Additionally, it serves as the conjugate prior for the likelihood which is defined as a binomial distribution, 
a primary choice for modeling a sequence of independent observations, each having a binary outcome. The main benefit of a conjugate prior is that it gives a closed-form expression for the posterior, significantly reducing the numerical burden. In general conjugate priors give a very intuitive insight into how a likelihood function updates a prior distribution. Lastly, the Beta distribution has the flexibility to capture a diverse spectrum of prior beliefs regarding the probability of success. We then update our beliefs about $\eta_{\text{obs}}$ based on the results of our survey of exoplanets in our sample for each $N \in[1,100]$.


The general definition of the beta distribution is:
\begin{equation}
    f(\eta_{\text{obs}} | \alpha, \beta) = \frac{\Gamma(\alpha + \beta)}{\Gamma(\alpha) \Gamma(\beta)} \eta_{\text{obs}}^{\alpha - 1} (1 - \eta_{\text{obs}})^{\beta - 1}
\end{equation}
 where $\Gamma$ denotes the Gamma function. In the following sections we explain how $\alpha$ and $\beta$ are defined in the particular context of our analysis.
From the practical perspective, we choose the conjugate Beta-Binomial model for its computational simplicity and ease of interpretation. It makes the calculations of Bayesian posteriors extremely straightforward \citep[for more details, e.g.][]{johnson2022bayes}, which
is described in Section \ref{sec:met_appl}.\\

\subsection{Defining priors}\label{def_priors}

Prior functions are used in Bayesian statistics to specify our belief in the possible distribution of a variable before performing the experiment. The prior can be  more or less informative, depending on the extent of previous knowledge it conveys regarding the variable. In situations where very limited or no a priori information is available, an ``uninformative'' prior is typically employed. It is essential to recognize that even in such cases, a small degree of objective information is factored into the analysis, such as the range within which the variable is defined.

\begin{figure}[h!]
    \centering
    \includegraphics[width=\textwidth]{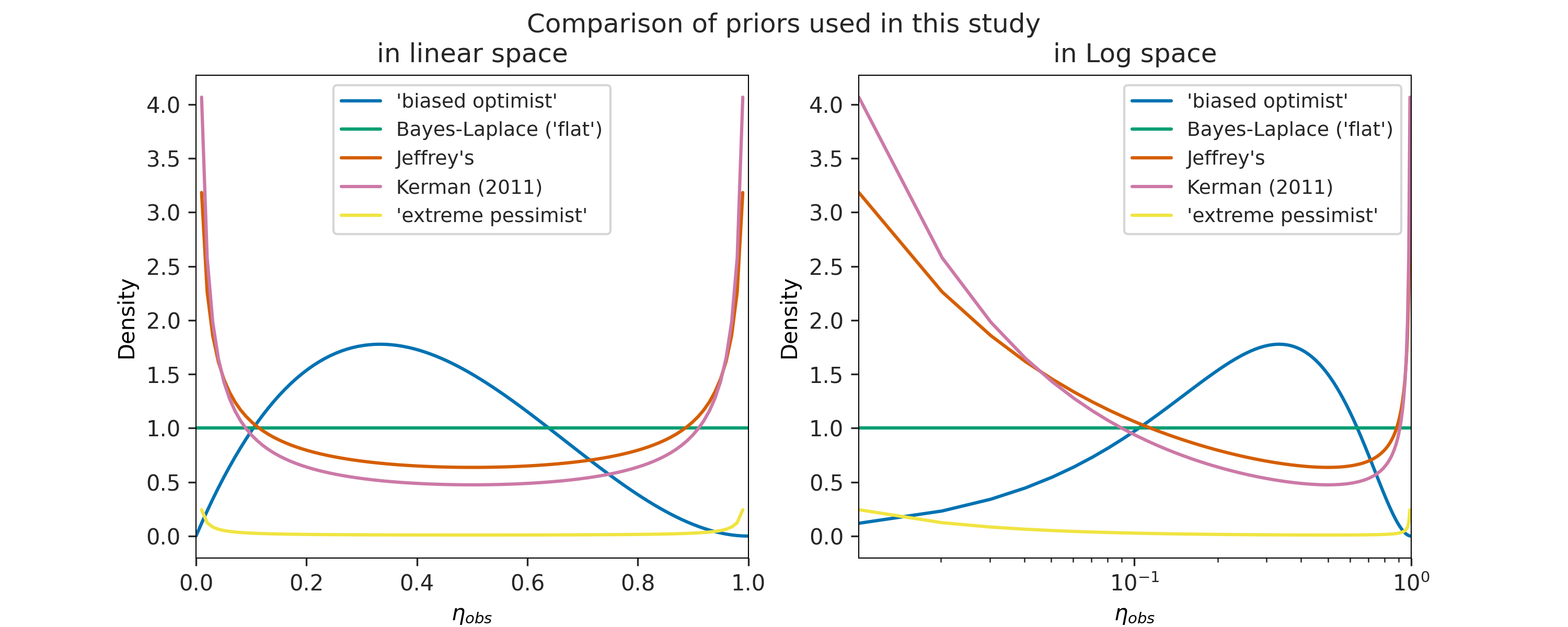}
    \caption{Comparison of the various priors used in this study. While the Bayes-Laplace (`flat') prior might seem ``uninformative'' when plotted in linear space, the logarithmic scale shows how it ``optimistically'' puts a lot of weight in values of $\eta_{\text{obs}}\sim 1$. This might not be the best choice for astrobiological questions, if one assumes that processes are either extremely rare or very frequent. Therefore we also chose more ``neutral'' priors such as the Jeffreys or Kerman ones, and also one extremely ``pessimistic'' one, giving more weight to $\eta_{\text{obs}}\approx 0$. }
    \label{figapp_priors}
\end{figure}

As we mentioned earlier, the Beta function can represent varying prior distributions depending on the choice of the parameters $\alpha$ and $\beta$. In our analysis, we adopted the following five choices of priors (see Figure \ref{figapp_priors}):

\begin{itemize}
    \item \textbf{Bayes-Laplace (``flat'') prior:} This prior assumes that all values of the variable are equally likely. It is given by the Beta distribution with $\alpha_{\text{flat}} = 1$ and $\beta_{\text{flat}} = 1$. Although this prior is sometimes considered ``uninformative'', it is in fact rather optimistic in our context, as it gives a large relative weight to values of $\eta_{\text{obs}}$ close to unity.\\

    \item \textbf{Jeffreys prior:} This prior is a more appropriate choice for the ``uninformative'' case, as it gives a similar weight to the extreme cases of $\eta_{\text{obs}}\approx 0$ and $\eta_{\text{obs}}\approx 1$. It is derived from the Fisher information matrix, and it is given by the Beta distribution with $\alpha_{\text{jeff}} = 0.5$ and $\beta_{\text{jeff}} = 0.5$.\\

    \item \textbf{Kerman prior \citep{10.1214/11-EJS648}:} This prior falls roughly in between the Jeffreys prior and the pessimistic prior and can be considered a ``neutral'' prior, because it is constructed so that the posterior distributions contain approximately 50 per cent probability that the true value is either smaller or larger than the maximum likelihood estimate.
    It is given by the Beta distribution with $\alpha_{\text{ker}} = 0.333$ and $\beta_{\text{ker}} = 0.333$.\\

 \item\textbf{Solar system ``biased'' optimist:} This prior is informative. It is given by the Beta distribution with $\alpha_{\text{opt}} = 2$ and $\beta_{\text{opt}} = 3$. This prior would for example be derived from a posterior after using a flat prior ($\alpha_{\text{flat}} = \beta_{\text{flat}}=1$) and the Solar System's sample of terrestrial, temperate (\textit{``optimistic habitable zone''}) planets as our baseline sample. In that case we would add 1 positive observation (Earth) to $\alpha$ and 2 negative observations (Mars, Venus) to $\beta$ when deriving the general \textit{"fraction of planets in the optimistic HZ that are habitable/inhabited"}. In that sense it would represent an interpretation suffering from selection bias (we can only exist as observers on a living planet) leading to a presumably too optimistic prior. While this prior is still relatively low in information/belief by only slightly preferring solutions around $\sim$0.3, it tends to produce rather ``optimistic'' posteriors (i.e., still believing in a relatively high fraction of living planets) in comparison to other more ``pessimistic'' priors (i.e., preferring solution with very low frequencies) when given the same set of updated information through additional non-detection observations (see also Appendix Figure \ref{figapp_priors2}.)\\
 
    \item \textbf{Extreme pessimistic prior:} With weights mainly close to 0 and 1 this prior is also very informative in the sense that it assumes that the given feature will either be found for all or none of the observations. This and similar priors (e.g., the classic Haldane prior)\footnote{We do not use the Haldane prior ($\alpha=\beta=0$) because it yields improper posterior distributions for the extreme outcomes such as $N$ or 0 positive observations in $N$ trials.  are discussed and used widely in  astrobiological reasoning, for example also in the context of the ``fine tuning problem'' \citep{2024arXiv240707097K}. Once presented with only a few negative results, the posteriors belief is that} the probability of success is very small, by giving a large relative weight to values of $\eta_{\text{obs}}$ close to 0. It is given by the Beta distribution with $\alpha_{\text{ext}} = 0.005$ and $\beta_{\text{ext}} = 0.005$.
\end{itemize}
 Figure \ref{figapp_priors2} in the Appendix exemplifies and illustrates why the flat prior can be considered optimistic: for one positive observation in a hundred tests, the resulting posterior is strongly peaked around 1\%, being quite confident that this is the actual fraction; conversely, for the extreme pessimistic prior, even for one in a hundred detection the posterior still assigns strong credence to very low fractions, $<10^{-4}$). In this context, the Jeffreys and Kerman prior can be considered ``neutral'', i.e. not too optimistic or pessimistic. 
 
\subsection{Simulation of observational scenarios}\label{sec:met_appl}

We use Bayesian inference to calculate $\eta_{\text{obs}}$, the fraction of planets possessing some binary observable feature, given a prior belief about this fraction and update our posteriors with increasing number $N$ of observations. Our prior belief is modeled by the above mentioned beta distributions with shape parameters $\alpha_{\text{prior}}$ and $\beta_{\text{prior}}$.

The posterior distribution for the fraction of  planets having the desired feature, after $n_{pos}$ positive cases have been found in $N$ observations, is then calculated as follows:

\begin{equation}
f(\eta_{\text{obs}} | \alpha_{\text{prior}}, \beta_{\text{prior}}, N, n_{pos}) = \frac{\Gamma(\alpha_{N} + \beta_{N)}}{\Gamma(\alpha_N) \Gamma(\beta_N)} \eta_{\text{obs}}^{\alpha_{N} - 1} (1 - \eta_{\text{obs}})^{\beta_{N} - 1}
\end{equation}

with $\alpha_N=\alpha_{\text{prior}}+n_{pos}$ and $\beta_N=\beta_{\text{prior}}+N-n_{pos}$.\\

In other words, to update our interpretation of the survey results we would add the number of positive observations to $\alpha$ and the number of non-detections to $\beta$ to derive the posteriors at that stage of the survey.


In the first part of our simulations, we assume a survey of `perfect' non-detection, i.e., $n_{pos}$=0 with no ambiguity. We then update the prior distribution varying the survey size $N$ between 1 and 100, leading to posteriors in the form of Beta($\alpha$, $\beta + N$).

\subsection{Modelling imperfect individual observations}\label{sec:met_imp}

In reality, each individual observation (asking \textit{does my planet have the feature or not?}) will have a degree of uncertainty.
In this section we want to quantify the extent of this uncertainty and its potential impact on our results and statistical power, particularly under worst-case scenarios. But what does it actually mean if someone is \textit{"80 percent sure they did not find water"} or concludes 
 that they \textit{"can exclude the presence of life with 3 $\sigma$?"} Can we even find parameterizations to quantify this?\\

Bayesian inference allows us to update our beliefs about a parameter in light of new data. When the Beta distribution is used as a conjugate prior for a binomial likelihood, the posterior distribution remains a Beta distribution, enabling straightforward updating as new data is observed. Typically, this Bayesian update involves adjusting the parameters of the Beta distribution based on the number of successes $(n_{\text{pos}})$ and the total number of observations $(N)$.
However, in the presence of uncertainties—such as sample misclassification or potential false negatives—the effective amount of information provided by the data may be reduced or altered. To account for this, we adjust the parameters of the Beta distribution to reflect the true information content of the data, rather than treating all observations as equally reliable. This adjustment is based on the concept of effective sample size \citep[ESS, see e.g.,][]{fsr189}, which quantifies the information that a sample provides about a population parameter, accounting for dependencies among observations, variability, and other sources of uncertainty. In our analysis, we adjust for sample uncertainty $(U_{\text{s}})$ and interpretation uncertainty $(U_{\text{i}})$.
Sample uncertainty $(U_{\text{s}})$ reflects the belief that some observed planets may not belong to the target sample. For example, if a mini-Neptune is mistakenly included in a sample of rocky planets, it should not influence the estimation of $(\eta_{\text{obs}})$, the fraction of rocky planets possessing the desired feature. To account for this, each observation is weighted by the factor $(1 - U_{\text{s}})$, where $(U_{\text{s}})$ represents the level of sample uncertainty. The posterior distribution is then updated as follows:
\begin{equation}
\eta_{\text{obs}} \mid \text{data} \sim \text{Beta}\left(\alpha + n_{\text{pos}}(1 - U_{\text{s}}), \beta + N(1 - U_{\text{s}}) - n_{\text{pos}}(1 - U_{\text{s}})\right)
\end{equation}
In cases where no positive detections are observed $(n_{\text{pos}} = 0)$, the posterior distribution is given by:
\begin{equation}
\eta_{\text{obs}} \mid \text{data} \sim \text{Beta}\left(\alpha, \beta + N(1 - U_{\text{s}})\right)
\end{equation}
This is analogous to using a weighted likelihood, where the effective sample size is reduced to account for the uncertainty in the relevance of the data. By doing so, we ensure that the posterior distribution accurately reflects the confidence we have in the sample's composition.
Interpretation uncertainty $(U_{\text{i}})$, on the other hand, addresses the possibility that some planets are incorrectly labeled as not having the desired feature, leading to false negatives. To address this, we adjust the posterior by adding a fraction $(N U_{\text{i}})$ of the total observations as "hidden" positives to the $(\alpha)$ parameter, while simultaneously reducing the number of true negatives added to $\beta$ by $(N U_{\text{i}})$. The posterior distribution is then updated as follows:
\begin{equation}
\eta_{\text{obs}} \mid \text{Data} \sim \text{Beta}\left(\alpha + N U_{\text{i}} + n_{\text{pos}}, \beta + N(1 - U_{\text{i}})\right)
\end{equation}
For cases with no positive detections ($n_{\text{pos}} = 0$), the posterior distribution becomes:
\begin{equation}
\eta_{\text{obs}} \mid \text{Data} \sim \text{Beta}\left(\alpha + N U_{\text{i}}, \beta + N(1 - U_{\text{i}})\right)
\end{equation}
This approach is akin to a robust Bayesian analysis, where the model accounts for potential errors in data classification. By incorporating "hidden" positives, we adjust for the possibility of false negatives, leading to a posterior distribution that more accurately reflects the underlying parameter $(\eta_{\text{obs}})$.
By  adjusting for sample misclassification and data interpretation errors, we reduce the likelihood of bias in the posterior estimates.

In a (maybe more intuitive) sense these two approaches represent two extreme possibilities: for the $U_{\text{s}}$ an uncertainty of 0.1 would mean that each (negative) observation only counts with a weigh of 0.9. For the $U_{\text{i}}$ scenario, however an uncertainty of 0.1 would mean that each (negative) observation only counts with a weigh of 0.9 and also add a 0.1 fraction of a positive observation to the interpretation.\\
As an illustration, let us consider a scenario where we conducted 50 observations without observing the desired feature. If we are confident in the outcome of each observation at the 80\% level, the most pessimistic assumption would be that we were wrong 20\% of the times (i.e. that $U_{\text{i}}$ was 0.2), corresponding, on average, to 10 false negatives in our sample. The most optimistic interpretation would be that in these 20\% of cases we could just not say anything (i.e. as if the planet wasn't even from our sample, $U_{\text{s}}=0.2$). Practical examples for these two cases could be: (1) a priori knowledge (e.g. from completeness studies of the survey) of a certain rate of wrongly labeled planets in the sample or (2) results from combined retrievals and (biological) interpretation that derive a certain rate of mislabeled planets that actually inhabit the desired feature. 

 We model both with three different levels of $U_{\text{s}}$ (50\%, 33\%, and 10\%)  and $U_{\text{i}}$ (20\%, 10\%, and 5\%) .


\subsection{Calculating limits}\label{sec:cal_lim}
We then calculate the upper and lower limits of the posterior distribution at each ``step'' of the survey, i.e. after each of the $N$ observations, by finding the values $f_{\text{obs,low}}$ and $f_{\text{obs,up}}$ such that:
\begin{equation}
\int_{0}^{\eta_{\text{obs,low}}} f(\eta_{\text{obs}} | \alpha_{\text{prior}}, \beta_{\text{prior}}, N,n_{pos})~d\eta_{\text{obs}} = 0.001,
\end{equation}
and
\begin{equation}
\int_{0}^{\eta_{\text{obs,up}}}f(\eta_{\text{obs}} | \alpha_{\text{prior}}, \beta_{\text{prior}}, N, n_{pos})~d\eta_{\text{obs}} = 0.999,
\end{equation}.

\begin{figure*}[t!]
    \centering
    \includegraphics[width=\textwidth]{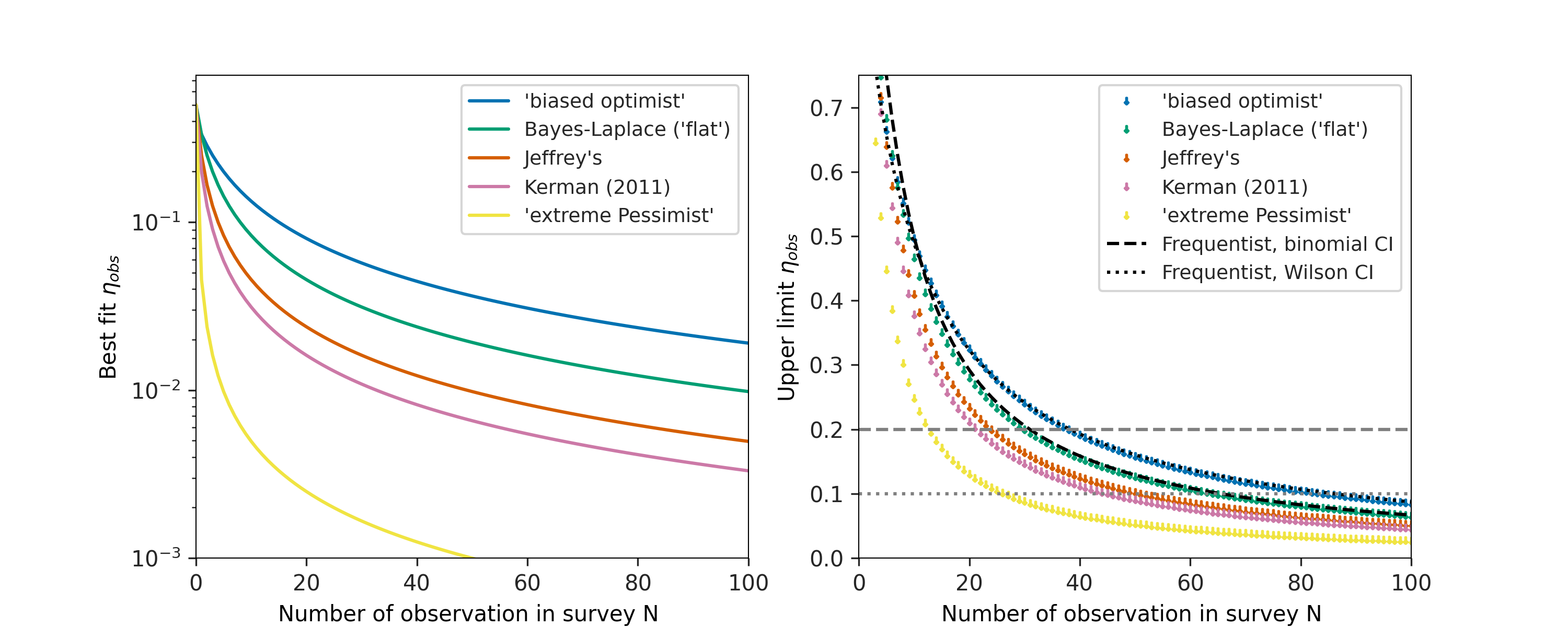}
    \caption{Final results for the case of a null-result survey of "perfect" observations. Left: Best fit from Bayesian analysis for the fraction $\eta_{\text{obs}}$ as a function of the number $N$ of non-detections. The Frequentist best fit corresponds to $\hat{\eta}_{MLE} = 0$, and hence can not be represented in this plot. Right: 99.9\% upper limits derived from the credible intervals of the posteriors for the different prior beliefs. Here we also compare Bayesian to Frequentist uncertainty assessments, by including both an exact binomial confidence interval, as well as a Wilson confidence interval based on a normal approximation. In order to exclude scenarios with $\eta_{\text{obs}}$ $>$ 0.2 and 0.1 with 99 percent belief, 
    we would need more than $\sim$ 40 or 80 individual, ``perfect'' observations, respectively. As expected and discussed in Section \ref{def_priors}, the ``optimistic'' priors produce posteriors with higher values whereas the ``pessimistic'' priors are at the lower end of the spectrum.}
    \label{fig:final_comp}
\end{figure*}

\subsection{Comparison to Frequentist Benchmark}\label{sec:met_freq}

We compare the results of the Bayesian approach with those of a Frequentist approach to provide a comprehensive analysis and ensure consistency across methods. It also allows us to show that our general conclusions are method independent. The key distinction of the Frequentist approach is that it does not incorporate prior beliefs; probabilities are solely derived from the observed event frequencies. Hence, let $n_{\text{pos}}$ represent the number of successes resulting from $N$ independent trials with unknown success probability $\eta$, where the random variable $n_{\text{pos}}$ follows a binomial distribution (as a repetition of Bernoulli trials):
\begin{equation}
n_{\text{pos}}\sim Bin(N,\eta)
\end{equation}
with the following likelihood:

\begin{equation}
L(\eta | N, n_{\text{pos}})=\binom{N}{n_{\text{pos}}}\eta^{n_{\text{pos}}} (1-\eta)^{(N-n_{\text{pos}})}.
\end{equation}

Due to the absence of observations of positive events ($x=0$), the maximum likelihood estimator of $\eta_{\text{obs}}$ 

\begin{equation}
\hat{\eta}_{mle} = \frac{n_{\text{pos}}}{N}
\end{equation}

will be 0. Thus, to overcome related computational shortcomings while building the confidence intervals, we employ an exact binomial confidence interval, with a lower limit at 0 and the upper limit calculated from the Cumulative Distribution Function of the Binomial based on a $99.9\%$ one-sided confidence interval. We also compare it with a conservative $99.9\%$ Wilson confidence interval based on the normal approximation, to ensure reliability of our results under larger sample sizes (for $N \geq 30$). Thus, the lower bound of the Wilson confidence interval is defined as $CI_{wilson_{L,0.999}}$ falls at 0. The upper bound is defined as $CI_{wilson_{U,0.999}}=min(1, CI^*_{wilson_{U,0.999}})$, with 

\begin{equation}
CI^*_{wilson_{U,0.999}} = \frac{1}{1 + \frac{z_{\alpha}^2}{N}}(\hat{\eta}_{mle} + \frac{z_{\alpha}^2}{2N} \pm \frac{z_{\alpha}}{2N}\sqrt{4N\hat{\eta}_{mle}(1 - \hat{\eta}_{mle}) + z_{\alpha}^2}).
\end{equation}

\section{Results}
\subsection{Observations with ``perfect'' individual non-detection}

The results for surveys with ``perfect''\footnote{i.e. no uncertainties in our sample population or observation interpretation} non-detection for each of the $N$ individual observations  are shown
in Figure \ref{fig:final_comp}. There we show the posterior mean (``best fit'') from the Bayesian analysis described above for the fraction $\eta_{\text{obs}}$ as a function of the number $N$ of non-detections, as well as the 99.9\% upper limits derived from the credible intervals of the posteriors, for the different prior beliefs. We chose 99.9\% as our threshold value to guide the intuition (particularly in public outreach or policy making contexts)  that ``there is only a 1 in 1000 chance'' that the \textit{real} number is \textit{not} below our derived upper limit. Also for an approximately normal data set, this threshold corresponds to a value slightly beyond three standard deviations (3 $\sigma$ $\sim$ 99.7\%). Around $N=20$ all priors ``agree'' on a ``best fit'' lower than $\sim$ 0.05, while after $\sim$40 observations they all ``agree'' on $\eta_{\text{obs}}<$0.15 with 99.9\% confidence.

In Fig.~\ref{fig:final_comp} (right, dashed/dotted) we also show a comparison with the results one would get using the Frequentist approaches to the problem. The confidence interval of the exact binomial aligns with the credible interval from the flat and uninformative prior, while the Wilson confidence interval aligns to the optimist credible interval for large sample regime. Although we plot those upper limits together with Bayesian credible intervals, it is important to note that confidence intervals and credible intervals encapsulate alternative meanings and probabilistic interpretations \citep{jin2017approximate}. Nevertheless, they are expected to converge in asymptotic behavior.

 In summary, we would need more than $\sim$ 40 or 80 individual, ``perfect'' observations, respectively, to exclude scenarios with $\eta_{\text{obs}}$ $>$ 0.2 and 0.1 with 99.9\% belief (or confidence in the Frequentist framework). 
 A similarly large number of observations is needed to reduce the impact of the chosen prior to be smaller than 5-10\% (see Figure \ref{fig:final_comp_diff}) on the final derived values.

\subsection{Modelling observational uncertainty in individual non-detection}

\begin{figure*}[t!]
    \centering
    \includegraphics[width=0.95\textwidth]{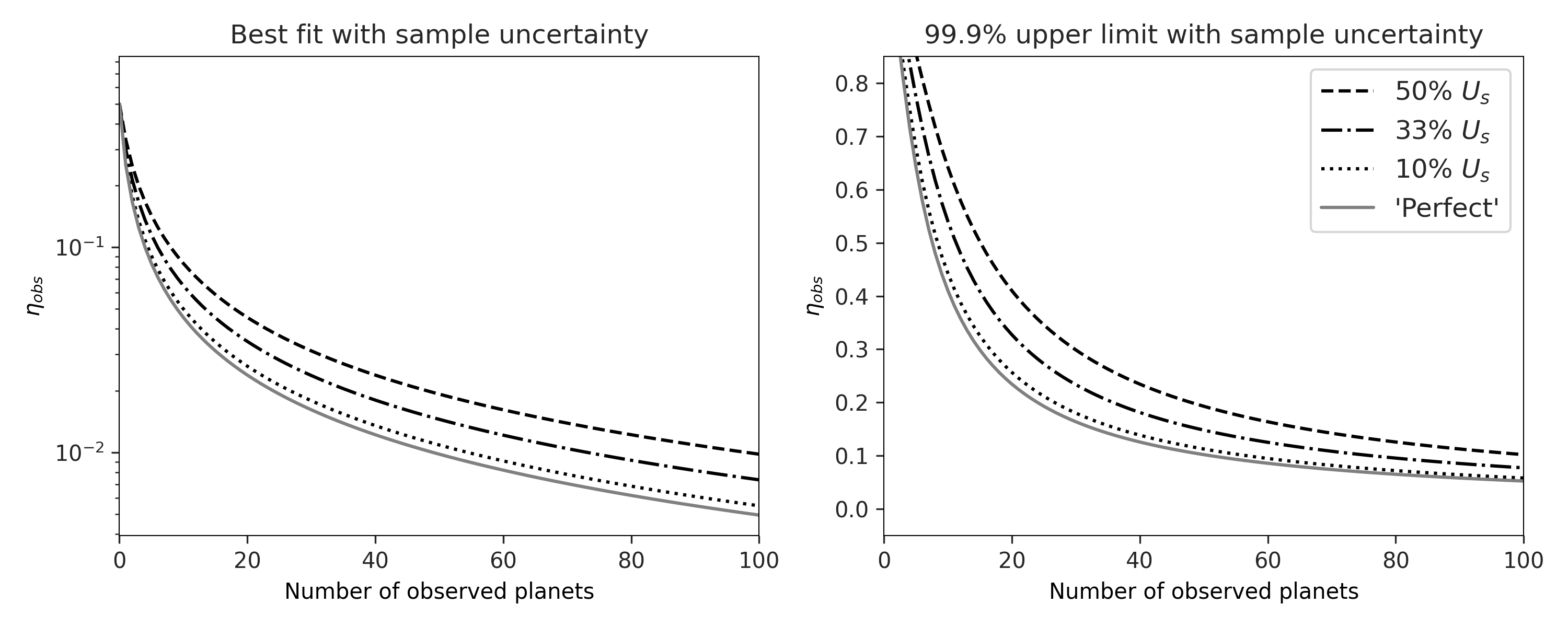}
    \includegraphics[width=0.95\textwidth]{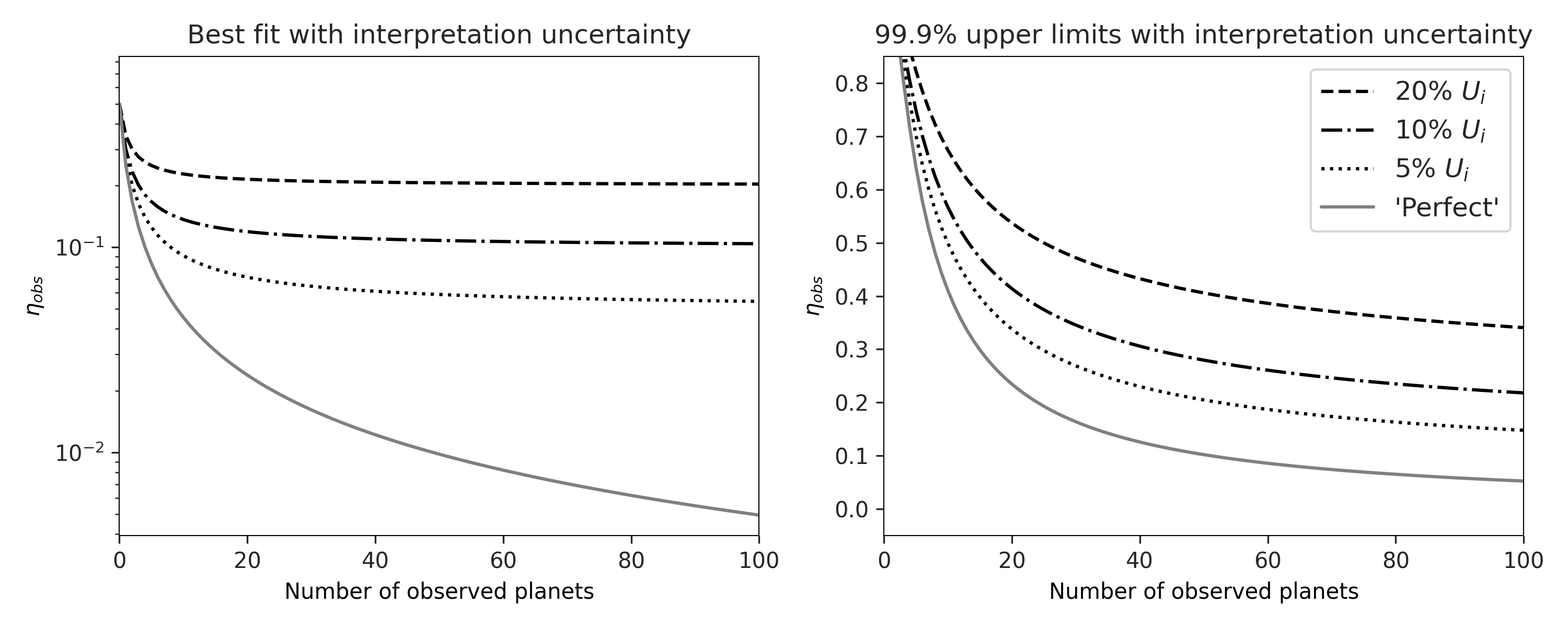}
    \caption{Testing different levels of "imperfect" individual observations, modeled as a sample (top) and interpretation uncertainty (bottom). Showing the difference between assuming "perfect" individual observations (solid line) and  more realistic results with various levels of sample and interpretation uncertainty.  For clarity we show only the calculations using  the Jeffrey's prior, as it seems to represent a good compromise between all the priors in the first general analysis (see e.g., Figure \ref{fig:final_comp}).}
    \label{fig:final_comp2}
\end{figure*}

Figure \ref{fig:final_comp2} shows the best fit (left) as well as the corresponding 99.9\% upper limits (right) as a function of the number of observations $N$ in our survey, for different sample (top) and interpretation (bottom) uncertainties using the Jeffreys prior. 
The figure shows how cases with non-perfect observations, assuming a certain interpretation uncertainty, always produce higher values for both the best fit and the upper limit. Ultimately, the derived values are limited by $U_{\text{i}}$. In other words, if we observe a series of non-detections and (cautiously) assume in our interpretation that all observations beyond our confidence levels are wrongly interpreted, we can never derive values lower than the $U_{\text{i}}$. For instance, if we perform 100 observations assuming a 20\% rate of false interpretations, our best fit will never be smaller than 0.2, and the 99.9\% upper limit will never be smaller than 0.4, no matter how much we increase $N$.
This highlights the fact that it is only possible to derive meaningful fractions $\eta_{\text{obs}}$ if the observable feature is unambiguous and can be confirmed or excluded with confidence levels higher than the desired 
assigned belief on the final result for $\eta_{\text{obs}}$. It also emphasizes the need to `ask the right questions', i.e., to not conduct surveys for which our confidence in the interpretation of the individual observational is very low. As shown in Figure \ref{fig:final_comp2}, an interpretation uncertainty of only 5\% has a worse impact on our ability to constrain an upper limit than a sample uncertainty of 50\%. This is why we did not model $U_{\text{i}}$ larger than 20\%.

For the cases of sample uncertainty $U_{\text{s}}$ the problem is less pronounced: in that case we can actually compensate for the out-of-sample planets in our survey by observing a larger number of targets.
In other words, even if we observe a series of non-detections and have to  \textit{a priori} belief that 50 percent of all observations do not actually fall into our defined sample, we can just double our sample to $2N$ to derive the same upper limits as an unbiased survey with a perfect sample.

\subsection{Application to future survey missions}

In this section we compare our previous findings with real life scenarios for future mission concepts, namely LIFE and HWO. For completeness we want to also mention adaptive optics supported high-dispersion spectroscopy that will be possible with the Planetary Camera and Spectrograph for the Extremely Large Telescope  \citep[PCS/ELT,][]{2021Msngr.182...38K}. A hypothetical 10-year survey - assuming best case scenarios - with ELT/PCS could detect modern Earth-like oxygen levels in the atmospheres of up to 19 Earth-like exoplanet candidates (EECs) around bright, nearby M stars within 20 parsecs \citep[][]{2024arXiv240511423H}.

\begin{figure*}[t!]
    \centering
    \includegraphics[width=0.95\textwidth]{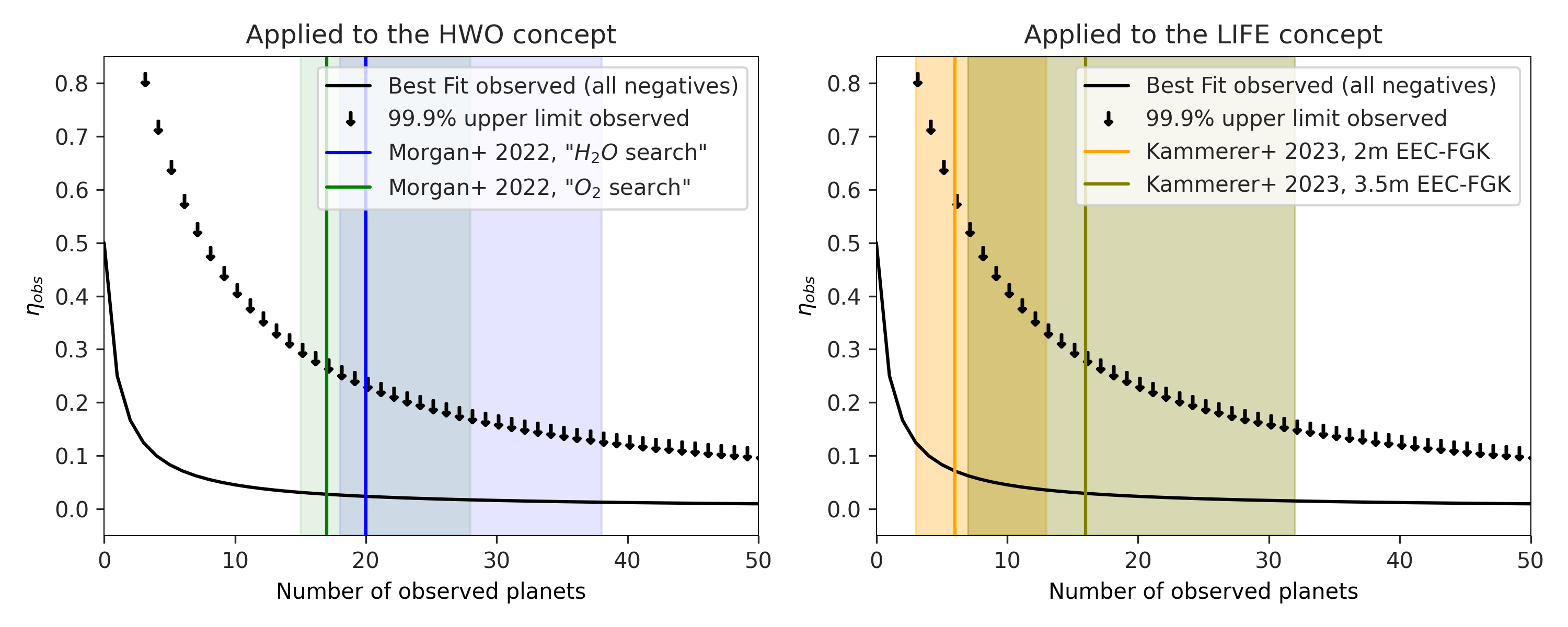}
 
    \caption{Comparison to potential survey sizes of the HWO mission and LIFE concept (using Jeffrey's prior). }
    \label{fig:survey_comp}
\end{figure*}
\subsubsection{The Large Interferometer for Exoplanets (\emph{LIFE})}

LIFE is a proposed space-based mid-infrared (MIR) nulling interferometer \citep{LIFE1}. This mission will be designed with the capability to investigate the atmospheric properties of a large sample of (primarily) terrestrial, temperate exoplanets. \citet{LIFE1}, \citet{LIFE2}, and \citet{LIFE6} discussed the detection yield of a LIFE-like mission. 
\cite{LIFE6} considered a 2 m aperture at 5\% throughput  baseline scenario for LIFE, which would detect (and eventually also characterize) $N_{2m/EEC} = 6_{-3}^{+7}$ exo-Earth candidates (EEC) around FKG stars. EEC here is defined as a planet with instellation flux range as received in the conservative HZ and an upper radius limit of 1.5 R$_{\oplus}$ but a lower limit of 0.8 $a^{-0.5}_{p}$ R$_{\oplus}$, where $a_p$ is the planet semi-major axis in units of AU \citep{2019JATIS...5b4009S}. Using an extended scenario with a 3.5-m aperture setup, \cite{LIFE6} derived $N_{3.5m/EEC} = 16_{-9}^{+16}$.

For the ``LIFE baseline'' $N_{2m/EEC} =6$ case, with clear (i.e. 100\% certain) non-detections of the observable feature the presented framework (using the Jeffreys prior) would derive: $\eta_{\text{obs}}=0.071$ and a 99.9\% upper limit of $\eta_{\text{obs}}<0.57$. This means that in case of an unsuccessful search for \textit{``life as we know it on an Earth twin''} in this sample of 6 planets we would (only) be able to constrain $\eta_{\text{Life}}$ for FGK stars to be smaller than 57\% with 99.9\% belief. 

For the second discussed ``LIFE extended'' $N_{3.5m/EEC}=16$ case, the framework presented here (again, using the Jeffreys prior) would derive $\eta_{\text{obs}}=0.029$ and a 99.9\% upper limit of $\eta_{\text{obs}}<0.28$. This means that in case of an unsuccessful search for habitable planets in this sample of 16 planets, LIFE would be able to constrain $\eta_{\text{Hab}}$ for FGK stars to be smaller than 28\% with 99.9\% belief.

In Appendix Section \ref{app:yielddisc} we compare different parameter spaces in radius and insulation also discuss LIFE's capability of detecting and characterizing terrestrial planets in the optimistic HZ (a wider criterion than EEC) of FGK stars.

\subsubsection{The Habitable Worlds Observatory (HWO)}

NASA's HWO\footnote{Based on the the Large UV/Optical/IR Surveyor \citep[\textit{LUVOIR,}][]{2019arXiv191206219T}, and the Habitable Exoplanet Observatory \citep[\textit{HabEx},][]{2020arXiv200106683G}} is a flagship-class mission proposed by the US Astro2020 decadal survey \citep{2021pdaa.book.....N} that  will utilize a combination of spectroscopic and imaging techniques to explore the atmospheres and surfaces of temperate and terrestrial exoplanets. For HWO we used numbers from \cite{2022SPIE12180E..20M}\footnote{Their upper and lower limit are derived by the extreme cases of (i) no knowledge about existing planets before the start of their search phase or (ii) perfect knowledge of all available target which allows them to skip the search phase.}: using a 6 m coronagraph scenario and constraining to ``exo-Earths''\footnote{ defined there as planets with  $R \in$ [0.8$\sqrt{a}$, 1.4] $R_\oplus$ orbiting at distances $a$ with an insulation equivalent to [0.95, 1.67] AU of the Sun, almost identical to the LIFE EEC criterion (see Appendix Section \ref{app:yielddisc})} around ``sun-like'' stars, they analyzed the yield of planets for which HWO could detect (or exclude) the presence of water or oxygene in the atmosphere based on a detection metric introduced there.

They show that they can measure a \textit{water observable metric} for $N_{H_2O} = 20_{-2}^{+18}$ and an \textit{oxygen observable metric} for $N_{O_2} = 17_{-2}^{+11}$ planets in the HWO survey. For $N_{H_2O}=17$, our framework would derive $\eta_{\text{obs}}=0.028$ and a 99.9\% upper limit of $\eta_{\text{obs}}<0.26$; for $N_{O_2}=20$ we get $\eta_{\text{obs}}=0.024$ and $\eta_{\text{obs}}<0.23$; as before, we are assuming that no positive detection was made in the sample. This means that in case of an unsuccessful search for oxygen in the HWO sample of 20 planets, we would be able to constrain $\eta_{O_2}$ for planets around FGK stars to be smaller than 23\% with 99.9\% belief, 
if none of the HWO observations will detect oxygen. Figure \ref{fig:survey_comp}  shows the derived values for the whole range of possible $N$.

\section{Discussion and Conclusions}\label{conclusions}

In this work we show that future surveys of exoplanets aiming at characterizing several dozens of targets are large enough to derive meaningful statistical results for the frequency of observable features, even from a series of non-detections. Furthermore we show, that at these sample sizes, the results inferred from such surveys will be only weakly dependent on the choice of priors in a Bayesian interpretation of the result and comparable to conclusions made in a Frequentist approach.

Nonetheless, it is essential to consider certain caveats and details in relation to these conclusions. First of all, we showed that our uncertainty in the individual interpretation of the observables has a strong impact on the 
belief in the final result, so that the uncertainties of individual observations must be very low in order to derive statistically significant frequencies.
Another important point is to clearly define the sample (and understanding its limitations) as sample uncertainties also weaken the final conclusions. This means that, for example, asking which is the \textit{``fraction of planets that have life''} will be an ill posed question for these purposes. Asking for \textit{``planets that have global biospheres''} might be slightly better suited, since it at least implies the chance for remote detectability; however, it still relies on a strong consensus within the scientific community about which remote signatures indicate unambiguous detection of life \citep[cf.][]{Catling_2018, fisher2023complex, 2022arXiv221014293M}. From an observational standpoint, asking for something of the sort \textit{``planets within certain radii and temperature limits that show methane and oxygen simultaneously above certain thresholds''} \citep[see e.g.,][]{2018AsBio..18..630M, 2014ApJ...792...90D} will be more feasible, in particular with regard to the ability to clearly say whether the feature was observed or not; framing the question in such a way, however, reduces the meaning and usefulness (in terms of generalizability) of the derived result. For example, how would we fit non-methane/oxygen biosignatures in this framework, if detected? How generally applicable is $\eta_{\text{Hab}}$ if potentially habitable sub-Neptunes are excluded from the sample? Similar arguments could be applied to the sample on which the results is obtained: besides the general caveat that the planetary systems in our Solar neighborhood are not necessarily representative of all the system in other galactic environments, and that observations are not necessarily independent (e.g., assuming there is no panspermia on these distances, or on multiple planets in the same system), biases in the sample itself (e.g., arbitrary cuts in stellar hosts) or the observation (e.g., it may be easier/clearer to detect habitability on certain types of planets) will limit the usefulness of the eventually derived fraction $\eta_{\text{obs}}$ in more general contexts.

 Despite the inherent challenges in estimating $\eta_{\text{Hab}}$ or even inhabited $\eta_{\text{Life}}$ with a small sample size, such as broader credible intervals and reduced precision, this still might offer several key useful aspects. First, preliminary insights gained from these estimates can serve as valuable guides for subsequent research directions in exoplanet studies, even if the data are limited. Establishing an initial baseline is crucial, as it provides a reference point for comparison with future, more comprehensive datasets. Moreover, tackling the complexities of small samples influences the development of sophisticated statistical methods.  Understanding the limitations of small sample data is also instrumental in informing resource allocation for future observational efforts. These early estimates, while statistically limited, can significantly influence research policies and funding decisions. We are currently working on a follow up paper that generalizes the study conducted here to all sorts of false negative and positive rates as well as underlying distributions of the observable feature to even further optimize future survey strategies. 

 Last but not least we want to remind the reader here that, even if this paper is about null-results, a single positive detection would be a watershed moment in humankind's history. 



\begin{acknowledgments}
This work has been carried out within the framework of the National Centre of Competence in Research PlanetS supported by the Swiss National Science Foundation under grant 51NF40\_205606. D.A. and S.P.Q. acknowledge the financial support of the SNSF.
A.B.K. acknowledges funding provided by the University of Belgrade - Faculty of Mathematics (contract 451-03-66/2024-03/200104) through grants by the Ministry of Science, Technological Development, and Innovation of the Republic of Serbia. Thank you to Geoffrey Johnson for his review and feedback during the drafting of this manuscript.
\end{acknowledgments}

%






\appendix

\section{Discussion of mission yields}\label{app:yielddisc}

\begin{figure*}[h!]
    \centering\includegraphics[width=\textwidth]{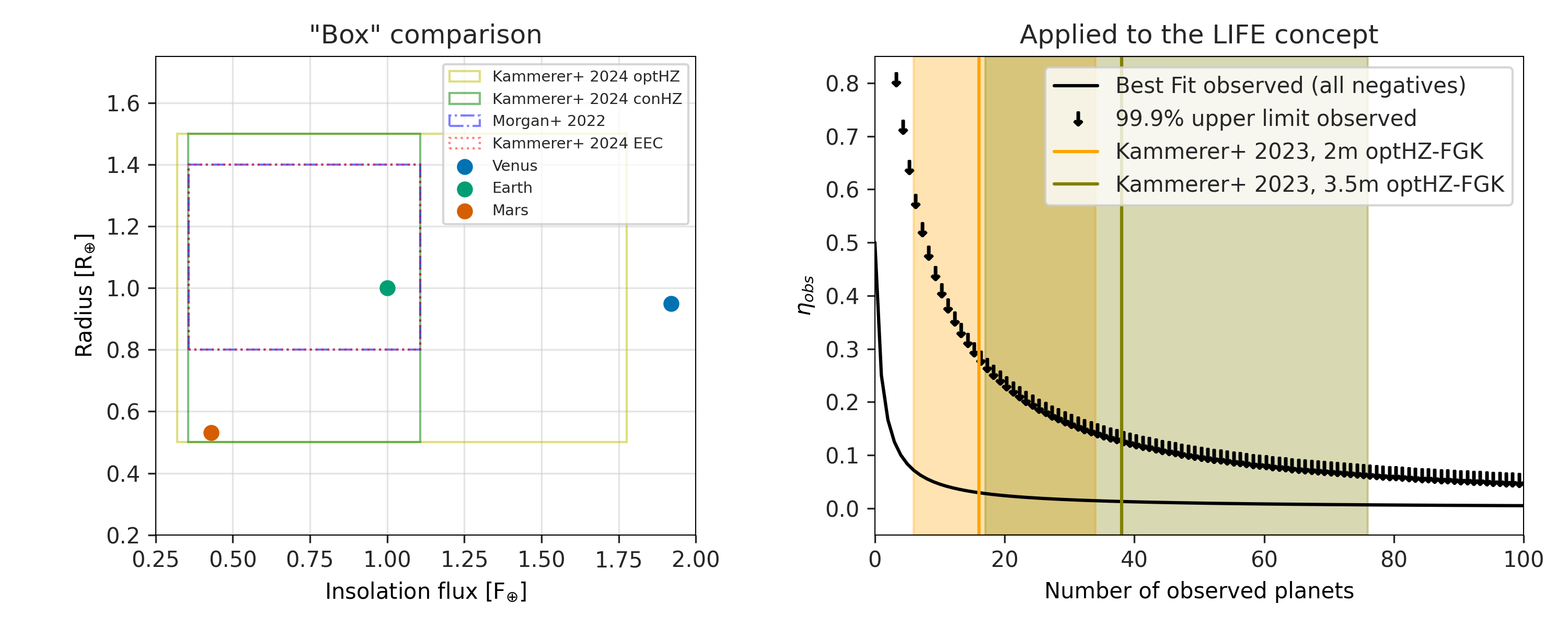}
    \caption{Left: comparison of the radius/insulation flux parameter spaces used in this paper to define the various exoplanet samples. Right: Same as Figure \ref{fig:survey_comp} for LIFE but with a wider "optimistic habitable zone (optHZ)" sample. }
    \label{fig:box_newLife}
\end{figure*}

In this appendix we give some more details on the expected yield of the future HWO and LIFE mission and explain the numbers used in the main paper.
Here we also note that \cite{LIFE6} and \cite{2022SPIE12180E..20M} used slightly different models of the underlying planet occurrence rate (see Figure \ref{fig:box_newLife}, which should be taken into account with regard to the comparison in Figure \ref{fig:survey_comp}.

\cite{LIFE6} using their ``2-m aperture at 5\% throughput  baseline scenario'' for LIFE, derive $N_{2m/optHZ} = 16_{-10}^{+18}$ detections of terrestrial planets in the optimistic HZ (a wider criterion than EEC) of FKG stars. Using an extended scenario with a 3.5-m aperture setup, \cite{LIFE6} this number grows to $N_{3.5m/optHZ} = 38_{-21}^{+38}$. This number is based on the most optimistic scenario for HZ rocky planet occurrence rates in \cite{LIFE6}.

For the second discussed ``LIFE extended'' $N_{3.5m/optHZ}=38$ case, the framework presented here (again, using the Jeffreys prior) would derive $\eta_{\text{obs}}=0.012$ and a 99.9\% upper limit of $\eta_{\text{obs}}<0.13$. This means that in case of an unsuccessful search for habitable planets in this sample of 38 planets, we would be able to constrain $\eta_{\text{Hab}}$ for FGK stars to be smaller than 13\% with 99.9\% belief.

\section{Illustrating the priors and difference between prior choices}\label{freq_comp}

\begin{figure*}[h!]
    \centering\includegraphics[width=\textwidth]{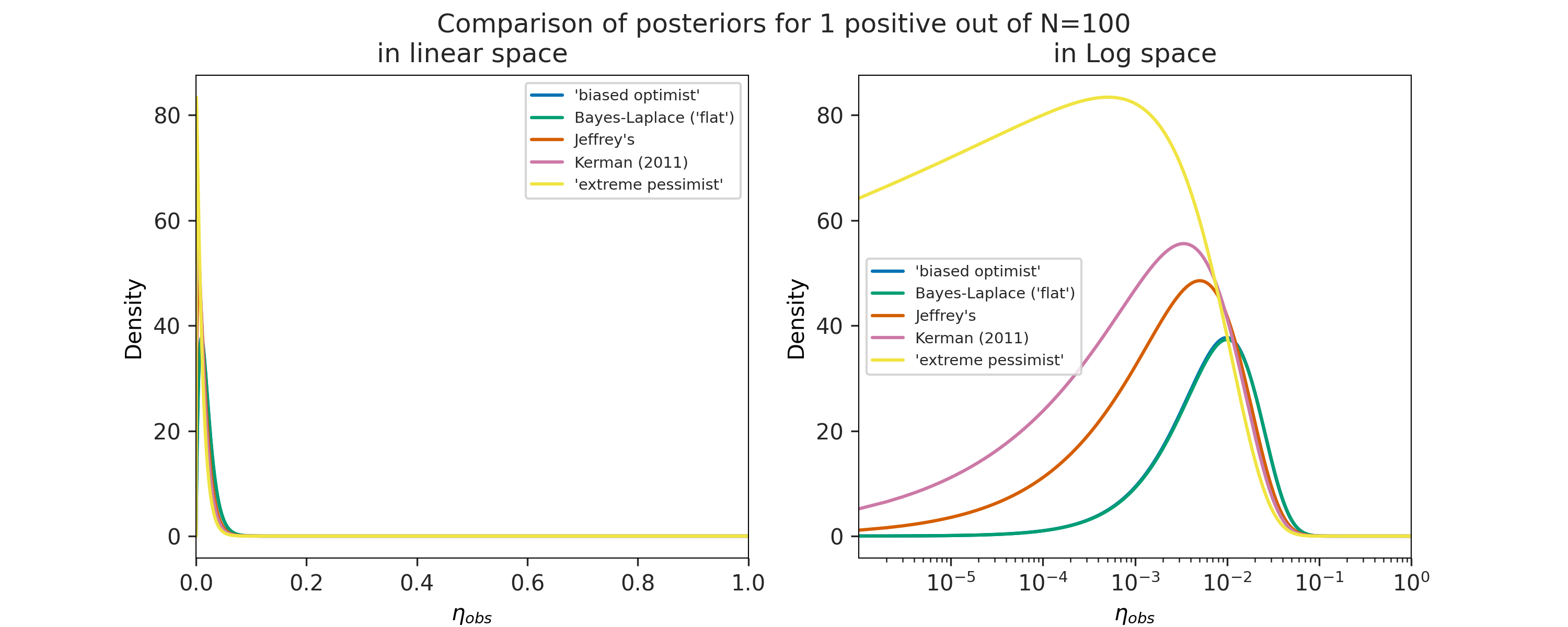}
    \caption{Example of the posterior distributions resulting from one positive detection in 100 (perfect) observations, assuming  different priors. While in linear space all priors seem to constrain the resulting value around 10$^{-2}$ (or 1\%), the logarithmic plot illustrates that the more ``pessimistic'' priors produces more statistical weight for values of $\eta_{\text{obs}}$ much smaller than 10$^{-2}$.}
    \label{figapp_priors2}
\end{figure*}

\begin{figure*}[h!]
    \centering
    \includegraphics[width=0.45\textwidth]{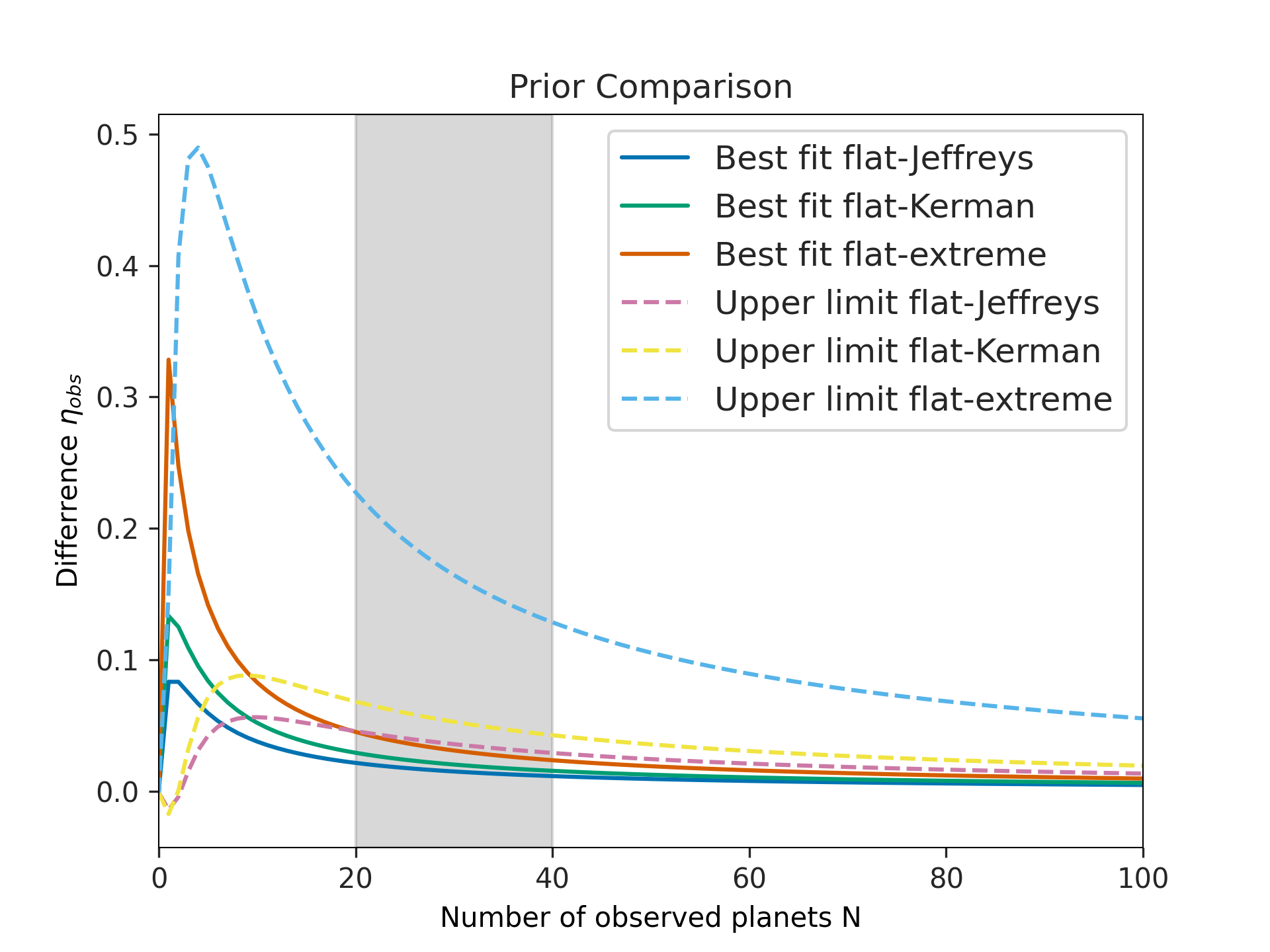}
    \caption{Comparison of results derived by this framework as function of the number of observations: differences in "best fit" and 99.9 percent upper limits between "optimistic" (flat) and "neutral/pessimistic" (Jeffreys/Kerman, extreme) prior analysis. N=20-40 seems to be enough to be (relatively) independent of the choice of priors.}
    \label{fig:final_comp_diff}
\end{figure*}

In this section we provide a few more figures to help the reader understand the use of the different priors in the 
context analysed here.
Fig. \ref{figapp_priors} compares the various priors used in this study.
Fig. \ref{figapp_priors2} illustrated the examplary case of the posterior distributions resulting from one positive detection in 100 (perfect) observations, assuming  different priors.
Fig. \ref{fig:final_comp_diff} shows the difference between the results of different priors as function of the survey size.

\bibliography{nullres}{}
\bibliographystyle{aasjournal}



\end{document}